\documentclass{elsart}

\usepackage{graphicx}

\input epsf

\begin{document}

\begin{frontmatter}

\title{The optical system of the H.E.S.S. imaging atmospheric Cherenkov 
telescopes \\
Part I: layout and components of the system}

\author[1]{K.~Bernl\"ohr}{,}
\author[3]{O.~Carrol}{,}
\author[2]{R.~Cornils}{,}
\author[1]{S.~Elfahem}{,}
\author[5]{P.~Espigat}{,}
\author[1]{S.~Gillessen}{,}
\author[2]{G.~Heinzelmann}{,}
\author[1]{G.~Hermann}{,}
\author[1]{W.~Hofmann}{,}
\author[1]{D.~Horns}{,}
\author[1]{I.~Jung}{,}
\author[1]{R.~Kankanyan}{,}
\author[1]{A.~Katona}{,}
\author[5]{B. Khelifi}{,}
\author[1]{H.~Krawczynski}{,}
\author[1]{M.~Panter}{,}
\author[5]{M.~Punch}{,}
\author[4]{S.~Rayner}{,}
\author[1]{G.~Rowell}{,}
\author[2]{M.~Tluczykont}{,}
\author[2]{R.~van~Staa}

\address[1]{Max-Planck-Institut f\"ur Kernphysik, P.O. Box 103980,
        D-69029 Heidelberg, Germany}
\address[2]{Universit\"at Hamburg, Inst. f\"ur Experimentalphysik,
       Luruper Chaussee 149, D-22761 Hamburg, Germany}
\address[5]{PCC College de France, 
       11 Place Marcelin Berthelot, Paris, France}
\address[3]{Dublin Institute for Advanced Studies, 
       5 Merrion Square, Dublin 2, Ireland}
\address[4]{Durham University, 
       The Observatory, Potters Bank, Durham, UK}

\begin{abstract}
H.E.S.S. -- the High Energy Stereoscopic System -- is a new system
of large imaging atmospheric Cherenkov telescopes, with 
about 100~m$^2$ mirror area for each of four telescopes, and 
photomultiplier cameras with a large field of view ($5^\circ$) and small pixels
($0.16^\circ$). The dish and reflector are designed to provide good
imaging properties over the full field of view, combined with
mechanical stability. The paper describes the design criteria
and specifications of the system, and the individual components
-- dish, mirrors, and Winston cones -- as well as their characteristics.
The optical performance of the telescope as a whole is the subject of a 
companion paper.
\end{abstract}

\end{frontmatter}



\section{Introduction}

Imaging atmospheric Cherenkov telescopes (IACTs) have emerged as the prime tool for
ground-based gamma-ray astrophysics in the VHE (very high energy) regime
\cite{weekes}. IACTs detect gamma-ray induced air showers via their
emission of Cherenkov light. The shower is imaged either by a single telescope
or simultaneously by several telescopes, which provide multiple views of 
an air shower and {which enable} 
the stereoscopic reconstruction of the shower geometry, resulting in 
improved angular resolution, energy resolution, and background suppression.
With their large effective detection areas of typically 50,000\,m$^2$ 
near threshold and a few 100,000\,m$^2$ at very high energy, IACTs are
sensitive to very small photon fluxes.
Observations with IACTs such as the Whipple~\cite{whipple}, 
CANGAROO~\cite{cangaroo}, Durham\cite{durham} and CAT~\cite{cat}
telescopes and with the HEGRA telescope system~\cite{hegra} have established 
about a
dozen cosmic sources of TeV gamma rays, both galactic and extragalactic
\cite{weekes}. The faintest source has a gamma-ray flux around 
$6 \cdot 10^{-13}$/cm$^2$s above 1 TeV, corresponding to about 3\% of the flux from the 
Crab Nebula, which is often used as a ``standard candle''. 

While the
results obtained in the last years have provided a great deal of 
information both about the acceleration mechanisms of high-energy particles
in astrophysical objects and about the propagation and interactions of
VHE photons, it is obvious that the sensitivity of current-generation instruments is
marginal in the sense that (a) only the strongest -- and often atypical --
sources of each class are observable, and that (b) the number of objects in
each source class is so small, that it is hard to decide which of the 
observed spectral features represent generic properties of the acceleration
mechanism, and which are simply consequences of the specific initial 
conditions or of the environment of a given source.

As a result, a number of next-generation IACTs and IACT systems are
under construction; as examples we mention the High Energy Stereoscopic
System (H.E.S.S.) with initially four telescopes located in 
Namibia~\cite{hess,hess-loi}, 
the CANGAROO four-telescope system in Australia~\cite{cangarooIII}, 
the VERITAS project consisting of seven telescopes in 
the USA~\cite{veritas},
and the single large MAGIC telescope
on the Canary Island of La Palma~\cite{magic}. 
Common goals of these instruments
are to lower the threshold for the detection of gamma-ray induced air showers
to 100 GeV or below, and to improve the sensitivity at higher 
energies by about one order of magnitude. Reflector areas range from
about 50\,m$^2$ for each of the CANGAROO telescopes to more than
230\,m$^2$ for the MAGIC dish.

All of these Cherenkov telescopes use optical systems consisting of
large tessellated reflectors, focusing the Cherenkov light onto photon
detectors made of bundled photomultiplier tubes (PMTs) to resolve the image
of the air shower.
With a characteristic size of air shower images of $1^\circ$ to
$2^\circ$, fields of view of the cameras range from $3^\circ$ to
$5^\circ$; at the lower limit, the camera is barely big enough to
contain shower images from a source located at the center of the field of view, 
whereas a large field of view {allows one to map sources} with a modest
spatial extension, such as galactic Supernova remnants. The size of
individual camera pixels (usually 0.1$^\circ$ to 0.2$^\circ$) is
matched to the scale of coarse features of the shower image. While
an even smaller pixel size {would provide further} improvements in
performance, a fine segmentation of the photon detector quickly
results in a very large number of pixels and in prohibitive costs.

Task of the optical system of IACTs is to collect the Cherenkov light
and to focus it onto the photocathodes of the PMTs. The transmission
of the optical system should be close to unity, to make optimum use of
the reflector area, and the point spread function should ideally be smaller than
the pixel size over the entire field of view. Given that data on 
a single source are often accumulated over 100 and more hours spread
over several months or even years, long-term stability of the optical
system is of crucial importance. While the imaging performance of
IACTs is modest compared to normal astronomical telescopes, the
cost-effective design of the dish and reflector of large Cherenkov telescopes
is non-trivial and represents a key ingredient for the quality of the observations
and the success of the instrument. The design and performance of
optical systems of Cherenkov telescopes are discussed, e.g.,
in \cite{whipple_opt} (Whipple), \cite{hegra_opt} (HEGRA) and 
\cite{cangaroo_opt} (CANGAROO).

With the goal both to present
the technical solutions for the various elements and to provide
a reference for future publications of physics results from H.E.S.S., 
this paper summarizes the design, implementation, and the components of the
optical system of the 
H.E.S.S. telescopes. 
The following chapters describe
the general design considerations, the reflector facets, the Winston cone light collectors
in front of the PMTs, and the mechanical layout of the dish and of the
mirror supports. The mirror alignment procedure and the 
optical performance of the telescope are the subject of a second
paper.

\section{Overview and design considerations}

This section outlines the general design considerations of the
optical system of the H.E.S.S. telescopes and the specifications
derived for its various components.

\subsection{The H.E.S.S. telescopes}

\begin{figure}
\begin{center}
\includegraphics[width=8.0cm]{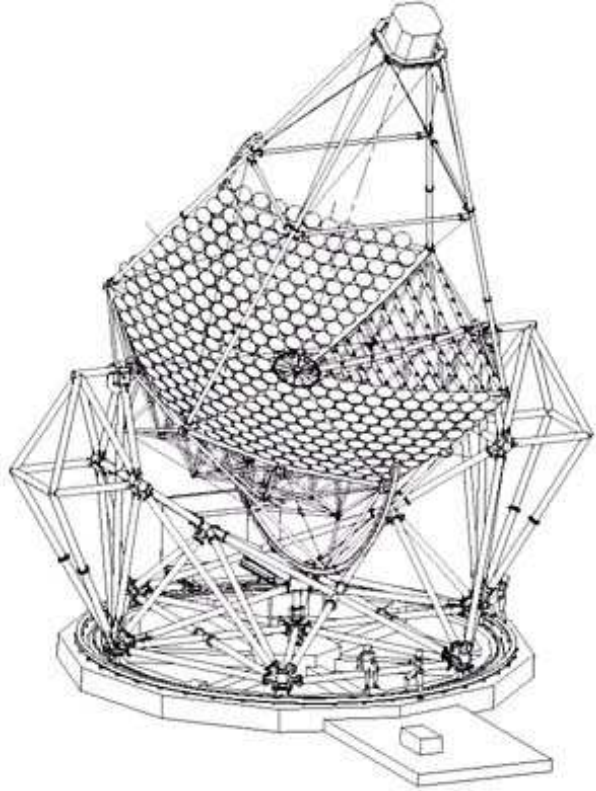}
\end{center}
\caption{A H.E.S.S. telescope, showing the steel space frame of
the dish and the telescope mount (the ``baseframe''). Mirrors are removed in one section
of the dish to view the support beams.}
\label{fig_telescope}
\vspace{0.5cm}
\end{figure}

Basic design criteria of the H.E.S.S. telescopes
\cite{hess,hess-loi}
(Fig.~\ref{fig_telescope}) were a reflector
area of about 100\,m$^2$, a camera field of view of $5^\circ$ diameter
and a pixel size of $0.16^\circ$, resulting in 960 pixels in each
camera. Given a typical yield of Cherenkov photons of about 
100/m$^2$\,TeV (in the 300 to 600 nm range) 
at about 2 km height a.s.l., combined with reflectivities of the mirror
of 80\% to 90\%, a similar transmission for the light-collection
funnels in front of the photomultiplier tubes, and an
average photomultiplier quantum efficiency around 15\%, this reflector
area guarantees a threshold around 100 GeV. At this energy,
images will contain
around 100 photoelectrons, and some detection capabilities should be 
provided
down to 50 GeV and below. The camera field of view of $5^\circ$
{allows one to observe} sources with an angular extension of a few 
degrees, and provides improved detection capabilities for distant
high-energy showers. The pixel size is matched to the size of 
image structures and results in a still manageable count rate 
induced by night-sky photons -- about 100\,MHz per pixel \cite{nsb}.
PMT signals are recorded by 1 GHz analog waveform recorders based
on the ARS ASIC; the entire electronics for triggering and signal
readout is contained in the approx. 900\,kg camera body \cite{punch_icrc}.

\begin{figure}
\begin{center}
\includegraphics[width=10.0cm]{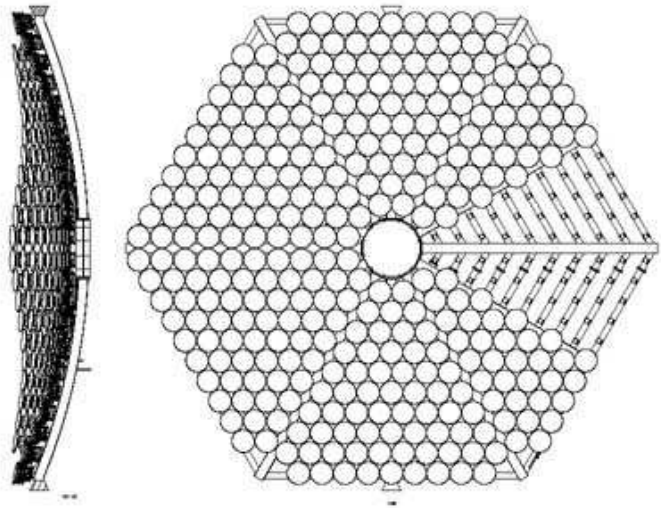}
\end{center}
\caption{Arrangement of mirror facets on the dish of a 
H.E.S.S. telescope; front view and cross section.}
\label{fig_mirr_arr}
\vspace{0.5cm}
\end{figure}

The reflector is segmented into 380 round mirror facets of 60 cm
diameter, supported by a steel spaceframe. 
The arrangement of mirrors is illustrated in Fig.~\ref{fig_mirr_arr}.
Motorized mirror
support units {are used for} the remote alignment of the mirror facets,
under the control of a CCD camera at the center of the dish,
viewing the photomultiplier camera. The alignment algorithm
is based on star images viewed on the closed camera lid. The
photomultiplier camera is supported by four arms
at the focal distance $f$ of 15\,m; combined with a characteristic
dish size $d$ of 13\,m the telescopes have $f/d \approx 1.2$.

The dish with the camera arms is mounted in alt-az fashion in
a rotating baseframe, supported at the ends of the elevation
axis by two towers. The baseframe rotates around a central 
bearing, on six wheels running on a 13.6\,m diameter rail.
Both in azimuth and elevation, the telescope is driven by
friction drives acting on special 15.0\,m diameter drive rails. The large lever
arm reduces the drive forces and the requirements on the reduction
gears. In both axes, the telescope slews at $100^\circ$/min. 
The position is sensed by shaft encoders attached at the 
central bearing and at one end of the elevation axis, with a 
digital step size of $10''$. A guide telescope mounted off-axis
on the dish and equipped with a CCD camera {should provide arc-second
pointing resolution}.

\subsection{Optics layout}

For cost reasons, the H.E.S.S. telescopes use -- like all other large
IACTs -- a segmented reflector composed of many individual mirror facets.
The facets are manufactured as spherical mirrors.
The possible options in the optics layout concern primarily the arrangement
of the mirror facets and the choice of the focal length for a given
reflector size. In addition, there is some freedom concerning the size and
shape of mirror facets. 

Classical arrangements of Cherenkov telescope reflectors follow either the Davies-Cotton layout
~\cite{davies-cotton},
or a parabolic layout. In the Davies-Cotton layout, generated originally
with solar concentrators in mind, all reflector facets have the same 
focal length $f$, identical to the focal length of the telescope as a whole.
The facets are arranged on a sphere of radius $f$. In a parabolic layout, 
mirrors are arranged on a paraboloid $z = r^2/(4 f)$, and the focal length
of the (usually spherical) mirror facets varies with the distance 
from the optical axis. 
While this adds a certain complication to the 
manufacturing process, the overhead is acceptable in particular when large
numbers of mirror facets of each focal length need to be manufactured. 
With small and perfect mirror facets, both approaches provide an
essentially point-like focus for rays parallel to the optical axis. 
Both suffer from significant aberrations for light incident at
an angle to the optical axis, the parabolic layout to a slightly larger
extent than the Davies-Cotton layout. Standard optics theory predicts that the
aberrations should increase linearly with the angle $\theta$ to the
optical axis, and with the inverse square of the ratio of focal length 
to reflector diameter, $f/d$. Another criterion is the time dispersion
introduced by the reflector, which should not exceed the intrinsic spread of the
Cherenkov wavefront of a few nanoseconds. Parabolic reflectors are isochronous --
apart from the minute effects introduced by the fact the individual
mirror facets are usually spherical rather than parabolic --
whereas the Davies-Cotton layout results in a spread of photon arrival
times at the camera; a plane incident wavefront results in photons
spread uniformly over an interval $\Delta T \approx d^2/(8 f c) {\bf\approx 5\,\mbox{ns}}$, with a rms
width $\Delta T/\sqrt{12} {\bf\approx 1.4\,\mbox{ns}}$.

\begin{figure}
\begin{center}
\includegraphics[width=5.5cm]{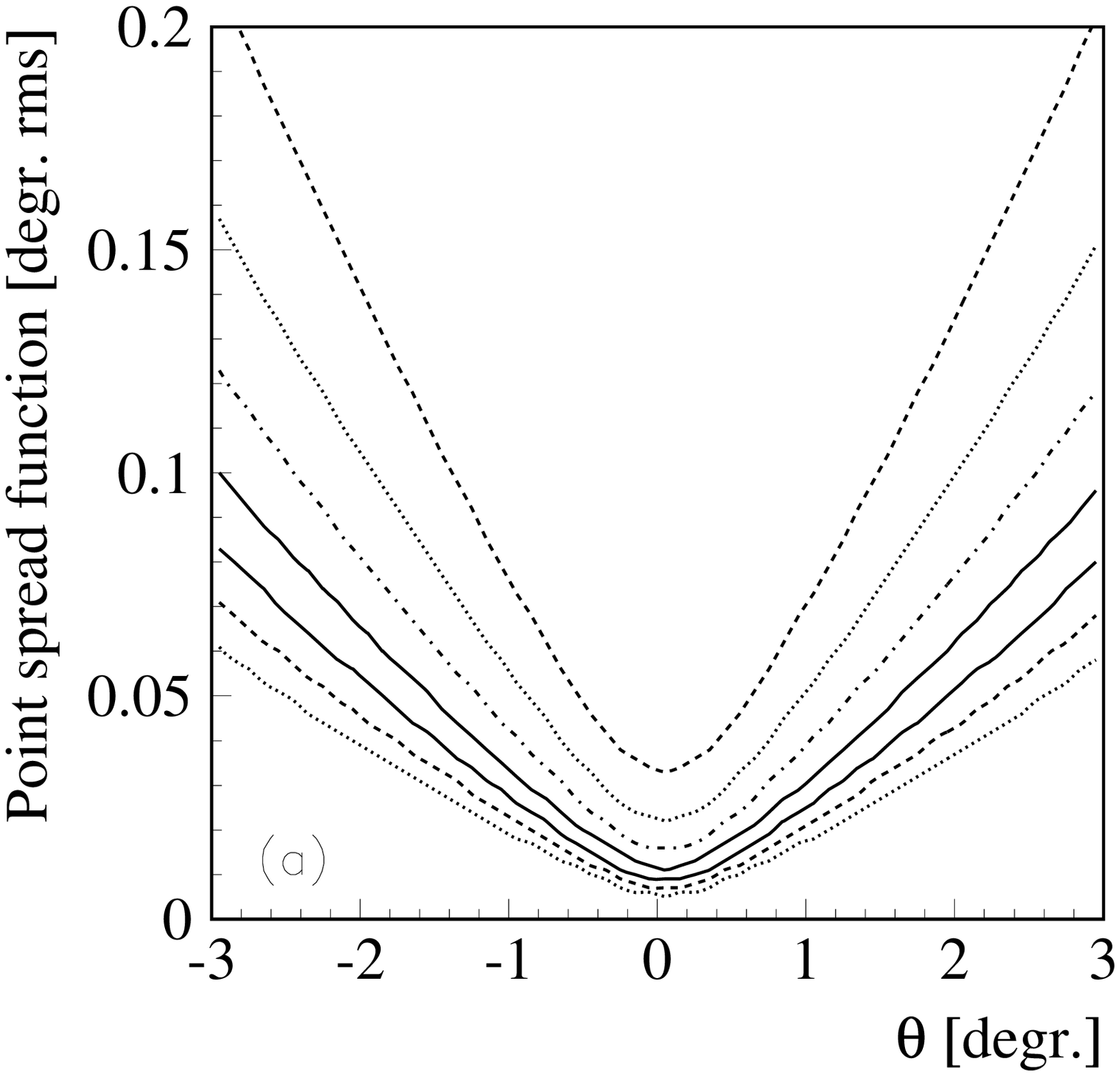}
\includegraphics[width=5.5cm]{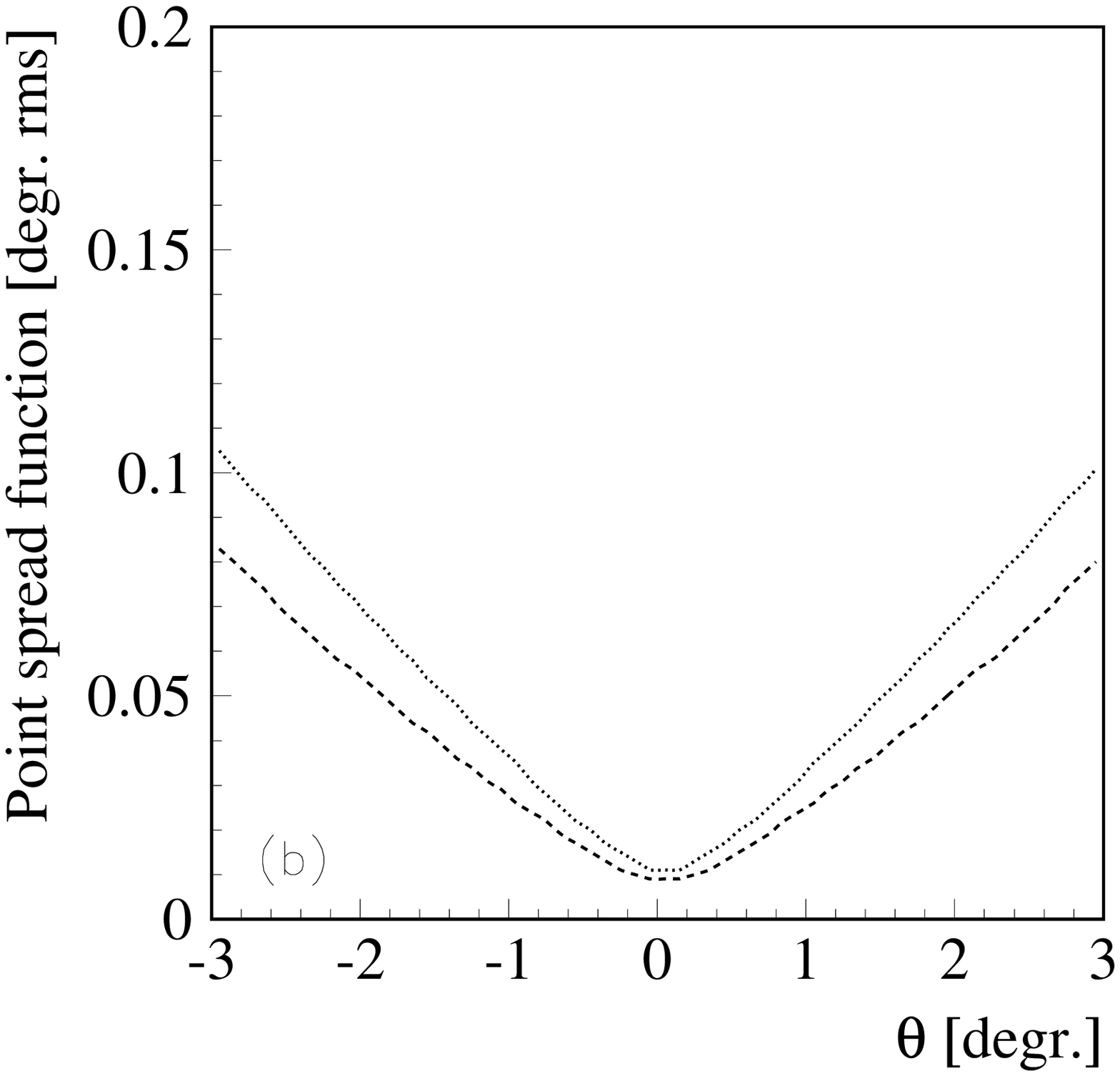}
\includegraphics[width=5.5cm]{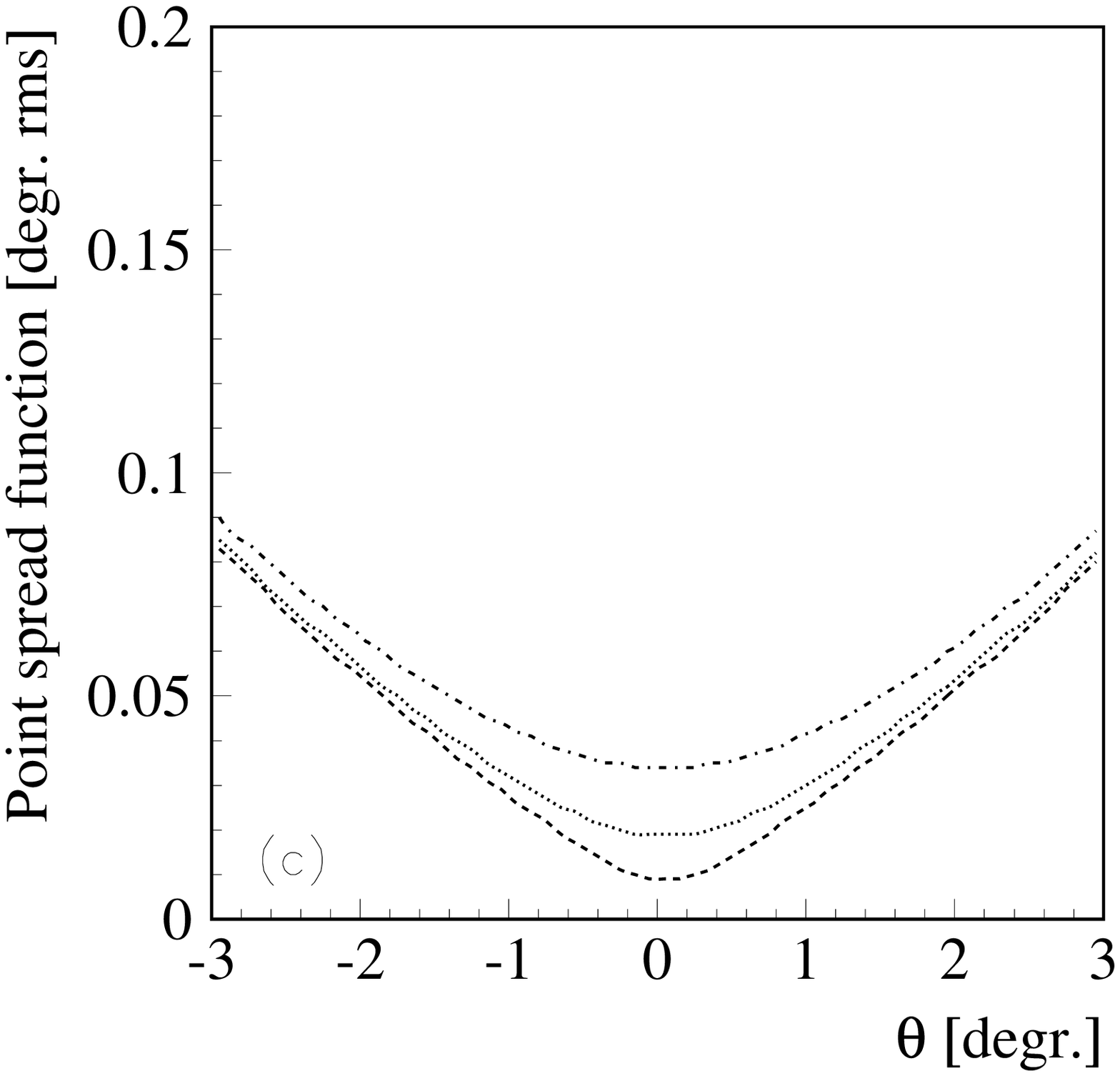}
\end{center}
\caption{Rms width of the point spread function -- i.e., the image
of a point light source at infinity -- as a function of the angle
$\theta$ to the optical axis of the telescope. The width is measured
in the radial direction; the width in the azimuthal direction shows
the same dependence, but is smaller by about 20\%.
(a) Comparison of Davies-Cotton layouts with
$f/d$ varying between 0.8 (top curve) and 1.4 (bottom curve) in steps of 0.1.
Imaging on axis is not perfect, due to the
assumed finite size of 60\,cm of the spherical mirror facets. 
(b) Comparison of a parabolic layout (top curve) and a Davies-Cotton layout
(bottom curve), for $f/d = 1.2$. 
(c) Influence of the point spread function $\sigma_0$ of the
individual mirror facets, for a Davies-Cotton layout with $f/d = 1.2$.
Top to bottom: mirror facets with 0.56, 0.28 and 0 mrad rms point spread
function, corresponding to 80\% of the light within 2, 1, and approx. 0\,mrad
diameter for a Gaussian distribution.}
\label{fig_im_1}
\vspace{0.5cm}
\end{figure}

The basic imaging characteristics are illustrated in 
Fig.~\ref{fig_im_1}(a)-(c), which are based on ray-tracing simulations
\cite{internal_ray_tracing,shadow,sim_hessarray} 
of different reflector geometries.
{The reflectors of the 
H.E.S.S. telescopes follow a 
Davies-Cotton layout. 
The time dispersion of such a reflector of $\approx 1.4$\,ns rms 
is still considered uncritical, since it is of the same order as the intrinsic time
spread of shower photons. The H.E.S.S.
signal recording electronics is designed to integrate
photomultiplier signals over 16\,ns so that signal are always fully contained, and 
the influence of the additional time smearing due to the mirror geometry is marginal.
The trigger electronics with its higher bandwidth would benefit from a parabolic
dish, providing slightly lower thresholds. On the other hand,
off-axis imaging of a Davies-Cotton mirror is slightly better 
(about 20\% in the spot size, see Fig.~\ref{fig_im_1}(b))
for the Davies-Cotton
layout, which is important since one of the design goals was the uniform response over the
large field of view, required for observations of extended sources. While the tradeoff
between the two variants is fairly balanced, the improved imaging led us to adopt the
Davies-Cotton layout.}

The choice of the $f/d$ ratio is a compromise
between the off-axis imaging quality and the increased cost, lower
resonant frequencies and mechanical complications caused by the 
need to support the camera at a larger distance from the dish. Under
these boundary conditions, $f/d \approx 1.2$ was adopted; with this
choice, the (rms) optical aberrations at $2^\circ$ off axis roughly
equal the (rms) pixel size.

The simulations shown in Fig.~\ref{fig_im_1}(a),(b) assume that the individual
mirror facets are perfect (spherical) mirrors. If the individual
facets are characterized by a point spread function of width $\sigma_0$
\footnote{Here and in the following, $\sigma_0$ denotes the rms 
width of the distribution after projection on one coordinate axis.
A corresponding two-dimensional Gaussian  point spread function is then
given by $\rho \sim \exp(-r^2/2 \sigma_0^2)$; the radii enclosing
68\% and 80\% of the light are given by $1.51 \sigma_0$ and
$1.79 \sigma_0$, resp.},
the overall point spread function should basically be given by a 
quadratic sum of the facet point spread function and the aberrations
caused by the dish geometry. The ray-tracing simulations of
Fig.~\ref{fig_im_1}(c) confirm that this is the case. A facet
point spread function of 0.3\,mrad rms will hardly be noticeably except 
right on-axis; a point spread function up to about 0.5\,mrad rms would be
tolerable, in comparison to the pixel size and the off-axis aberrations.
After discussions with manufacturers of mirror facets, the specification
was adopted that a mirror facet should image 80\% of the light within
a circle of 1 mrad diameter; for a Gaussian point spread function, this 
is equivalent to a rms 
width of 0.28 mrad. This specification is well below the critical performance,
and was accepted by the manufacturers to be reachable without significant extra
cost. Indeed, the quality control measurements discussed below show that
most mirrors pass this specification by a significant margin.

In addition to their quality, the size of the individual mirror facets
influences the optical performance; large (spherical) facets will
introduce additional optical errors. For the 60 cm facets used in H.E.S.S.
-- primarily for cost reasons, see below -- the resulting degradation is minimal.

Note that in all examples shown in Fig.~\ref{fig_im_1}, the mirror facets are aligned for
optimal on-axis imaging at a focal length $f$. 
Given the goal of a uniform imaging over a large
field of view, one might ask if other alignment schemes are possible
which improve the off-axis response at the expense of on-axis imaging,
or if mirror arrangements other than parabolic or Davies-Cotton provide
improved performance over the whole field of view.
Indeed, modest improvements of the off-axis response can be obtained
by slightly displacing the focal plane towards the reflector, by using
a curved focal plane, or by modifying the Davies-Cotton layout by moving
the outer mirrors closer to the focal plane, with a corresponding
change in their focal lengths~\cite{hess-loi}. However, in all cases
studied the improvements were small, and did not warrant the additional
complications or the degradation of the on-axis response.

As an additional option, systems with secondary reflectors were considered,
{which provide a partial compensation of aberrations}. However,
apart from the non-trivial construction and alignment of the secondary
reflector - which would also have to be realized as a segmented mirror --
such designs resulted in large effective focal lengths and hence
in much larger cameras (for a given field of view), with significant 
implications on the cost and performance of the photomultiplier tubes.

\subsection{System specifications and error budget}

The effective optical quality of the telescope is determined by a
number of factors, in particular
\begin{enumerate}
\item the aberrations inherent in the optics layout,
\item the optical quality of the individual mirror facets,
\item the tolerances in the design and fabrication of the 
dish, i.e., the precision with which
the mirror facets are positioned, 
relative to their nominal locations,
\item the quality
of the alignment of the facets, and
\item the deformations of the support structure under the influence of
wind loads, gravity load, temperature variations etc.
\end{enumerate}
A pixel of the H.E.S.S. cameras covers a rms projected angular range of about 0.7 mrad.
The point spread function of the reflector should not exceed this width
significantly, 
and ideally be smaller, in order to make full use of the camera granularity.
In the design of the H.E.S.S. telescopes, roughly equal contributions of optical
errors were allocated to the mirror facets (0.28\,mrad rms, in each projection), to the deformation of the
support structure under gravity and wind loads, temperature variations etc., and to the 
precision of the mirror supports and of the mirror alignment. A limit of 0.28\,mrad
rms on the influence of the carrier structure means that the structure has to be stable
to 0.14\,mrad rms, since errors are doubled due to the reflection of light. The same is
true for the mirror supports and the mirror alignment. Compared to the angular alignment
of mirror facets, their positioning turns out to be completely uncritical. Mirror facets
can be displaced along the optical axis by 10\,cm or more without significant impact. If all
components perform at the limit, the resulting spot size on axis is about 0.5\,mrad rms.

\section{Mirror facets}

In the initial stages of the design, a number of different 
mirror options were explored, including
\begin{itemize}
\item diamond-turned aluminum mirrors in the 50-60 cm diameter range,
\item aluminized ground-glass mirrors in the 50-120 cm diameter range,
\item mirrors consisting of a carbon fiber substrate covered by a
reflecting foil, and
\item membrane mirrors consisting of a steel membrane formed by slight
underpressure, covered with thin-glass mirrors or mirror foils, in the
120-240 cm diameter range \cite{membrane_mirror}.
\end{itemize}
With the test setup~\cite{dipl_elfahem} described below, 
glass mirrors from three different
manufacturers were tested, aluminum mirrors from two suppliers, and a
prototype membrane mirror built in-house.
While the H.E.S.S. Letter of Intent \cite{hess-loi}
still gave aluminum mirrors as the baseline option, 
the tests showed that the glass mirrors were consistently superior
to the aluminum mirrors available to us at the time, in particular in terms of
specular reflectivity. Concerning the spot sizes, the aluminum mirrors
were marginally acceptable. The membrane mirror showed poor image quality;
this option was abandoned because of the anticipated long time scale for further
development. 

Since the cost of glass mirrors was at the same level or below that of
aluminum mirrors, since weight was not of primary concern (and in any
case not much different between a glass mirror and a solid aluminum mirror),
and since glass mirrors had demonstrated their long-term stability in
the HEGRA and CAT telescopes, glass mirrors were chosen.

Among various mirrors sizes quoted, 
there was a clear cost optimum per unit mirror area
for facet diameters around 50\,cm to 60\,cm. Concerning facet shape, square and
hexagonal mirror facets {provide complete coverage of} the mirror area,
whereas round mirror facets will cause a 10\% loss for the same 
dish diameter. However, hexagonal or square mirrors
would have to be cut out of round mirrors, at significantly increased
cost per area. Therefore, round mirrors were chosen.

The resulting specifications for the mirror facets are
given in Table~1.

\begin{table}
\begin{center}
\begin{tabular}{|l|l|}
\hline
Material              & Aluminized optical glass, thickness $\approx 15$\,mm  \\
Protection            & Quartz coating for outdoor use\\
Diameter of mirror    & 600\,$\pm$\,1\,mm \\
Diameter of reflecting surface & $600^{+1}_{-2}$\,mm\\
Mounting              & Glued to support unit at three points \\
Focal length          & 15.00\,$\pm$\,0.25\,m \\
Point spread function & 80\% of light in 1\,mrad diameter \\
Specular reflectivity & at least 80\% between 300 and 600\,nm \\
\hline
\end{tabular}
\label{mirror_spec}
\vspace{0.3cm}
\caption{Initial specifications for mirror facets.}
\end{center}
\vspace{0.5cm}
\end{table}

\subsection{Mirror test setup}

\begin{figure}
\begin{center}
\mbox{
\includegraphics[width=6.0cm]{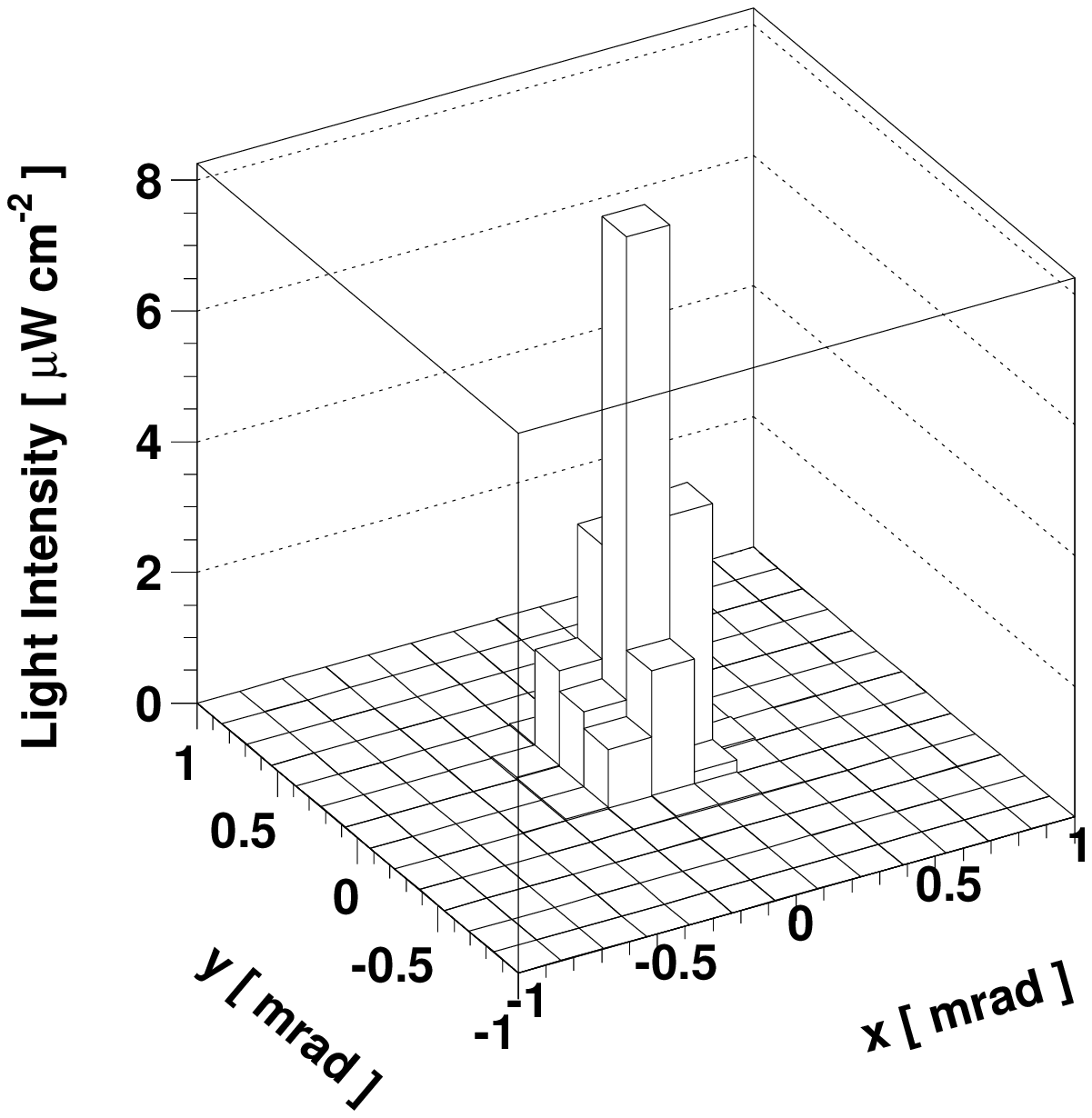}
\hspace{2cm}
\includegraphics[width=6.0cm]{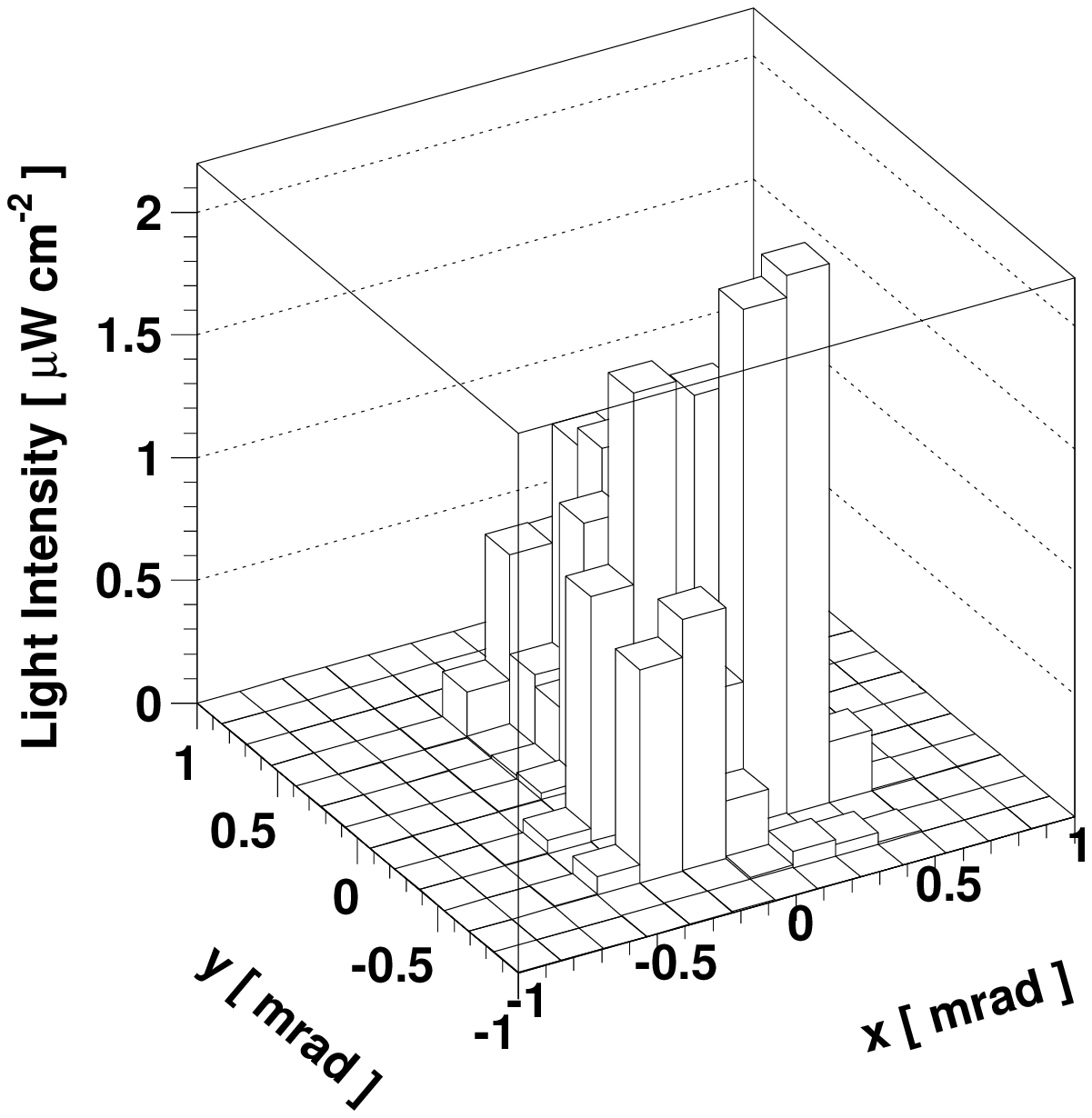}}
\end{center}
\caption{Examples for two scanned images: (left) a typical mirror,
and (right) a marginal mirror with a 1 mrad spot diameter.}
\label{fig_scans}
\vspace{0.5cm}
\end{figure}

In order to characterize the imaging properties and reflectivity
of a mirror, the mirror is placed at twice the nominal focal
distance (30.00\,m) from a light source equipped with a 3\,mm diameter
diaphragm. The light source illuminates the mirror uniformly;
over the 60\,cm diameter, variations in intensity are below 6\%.
The mirror is oriented such that the light source is
slightly off the optical axis, and a computer-controlled 3-axis
positioner carrying a photo diode is used to scan the resulting image. 
The resolution of the scan, about 0.33\,mrad FWHM, is 
determined by
a 10\,mm diaphragm in front of the scanning diode. Scans are 
carried out with a step size of 5\,mm, equivalent to 0.17\,mrad.
Fig.~\ref{fig_scans} shows two examples of scanned images. Integrating
outward from the center of gravity of the image, the diameter of
a circle enclosing 80\% of the reflected light is determined and 
is used to characterize the mirror.

In initial tests, images were scanned at different distances from the
mirror in order to determine the 
optical focus and the focal length; in the later series
tests, only the image at the nominal focal length is scanned. The 
measured spot width combines both focusing imperfections of the 
mirror and deviations from the nominal focal length; only this
quantity is relevant for our application.

In order to calibrate the light source, the scanning diode is moved
into the beam of the light source, such that it covers roughly the same
solid angle as the mirror, and the light flux per solid angle is
measured, given the known distance to the source and the
active area of the diode. By integrating the reflected light
over the scanned image and comparing its intensity
with the light incident on the mirror, the absolute specular reflectivity
is determined. 
The amount of light incident on the mirror is determined from the measured flux
per solid angle and the distance, assuming a 60\,cm diameter mirror. If the 
aluminized area of a mirror is smaller -- specifications require at least 59.8\,cm --
this will result in a slightly lower average reflectivity.
During the calibration and the measurements,
a second photo diode is placed off-axis in front of the
light source, such that it does not obscure the mirror, but is still 
in the region of uniform illumination. This second diode is used to
correct intensity variations of the light source during measurements.

A computer-controlled
filter wheel in front of the scanning diode enables wavelength-selective
measurements; four filters with 300, 400, 470 and 600 nm center 
wavelength and a bandwidth of about 10 nm are employed. We note
that this measurement determines the specular reflectivity under
realistic conditions, rather than the (frequently quoted by manufacturers, but
irrelevant) total reflectivity including diffuse reflection.

The accuracy of the reflectivity measurement was checked by
cross calibration with other methods and by repeated measurements
of reference mirrors over longer periods. We estimate the systematic
errors in reflectivity at 2.8\% at 300\,nm, and about 2\%
at 400, 470 and 600\,nm.

\subsection{Mirror characteristics}

Two suppliers were chosen for the mirrors of the H.E.S.S. telescopes,
COMPAS in Turnov (Czech Republic) for 1150 mirrors
and Galaktica in Yerevan (Armenia) for 600 mirrors.
Each mirror delivered was inspected visually for scratches or mechanical
defects, and the mirror diameter and the diameter of the reflecting
surface were measured. Finally, the reflectivity and spot size were
determined at the four wavelengths\footnote{The first batches were
measured at only two wavelengths, 310 and 470\,nm.}. At the time of 
this writing, all mirrors have been
delivered, and about 90\% of them have been measured
\cite{int_mirror1,int_mirror2}.

\begin{figure}
\begin{center}
\includegraphics[width=8.0cm]{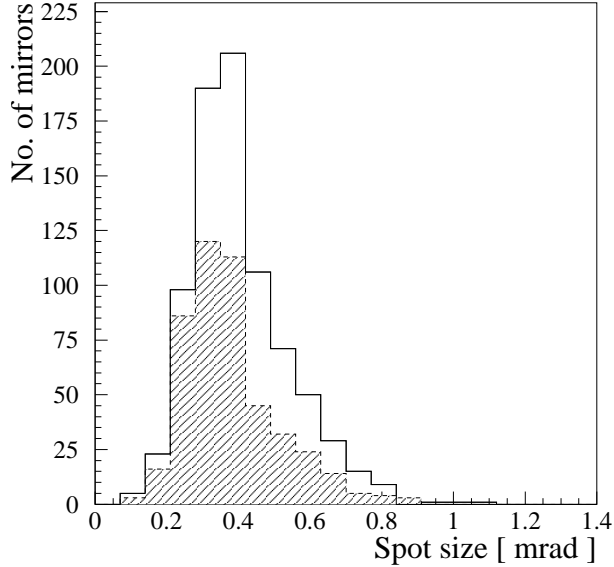}
\end{center}
\caption{Distribution of measured spot sizes, for the COMPAS
mirrors (full line) and the Galaktica mirrors (filled area), at 400 nm wavelength.
Shown is the diameter of a circle containing 80\% of the reflected
light. Due to size of the diaphragm in front of the
scanning diode, very small spot sizes (below 0.3 mrad) tend to be overestimated.}
\label{fig_angres}
\vspace{0.5cm}
\end{figure}

\begin{figure}
\begin{center}
\includegraphics[width=8.0cm]{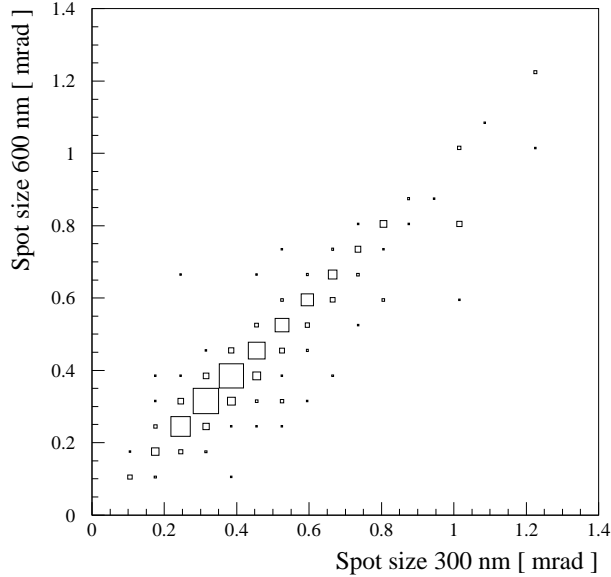}
\end{center}
\caption{Correlation between the spot sizes measured at 300 nm and 
600 nm.}
\label{fig_angcorr}
\vspace{0.5cm}
\end{figure}

Fig.~\ref{fig_angres} shows the distribution of measured spot sizes
for the COMPAS and the Galaktica mirrors, at 400 nm. Shown are all
mirrors, including those later rejected because of insufficient 
reflectivity. Most mirrors show a spot size around or below 0.5 mrad (diameter
for 80\% containment), a factor 2 below specifications. 
The spot sizes measured for a given mirror at different wavelengths
are highly correlated (Fig.~\ref{fig_angcorr}), as expected since 
the spot size is determined by the mechanical quality of 
the mirror shape and surface.

\begin{figure}
\begin{center}
\includegraphics[width=8.0cm]{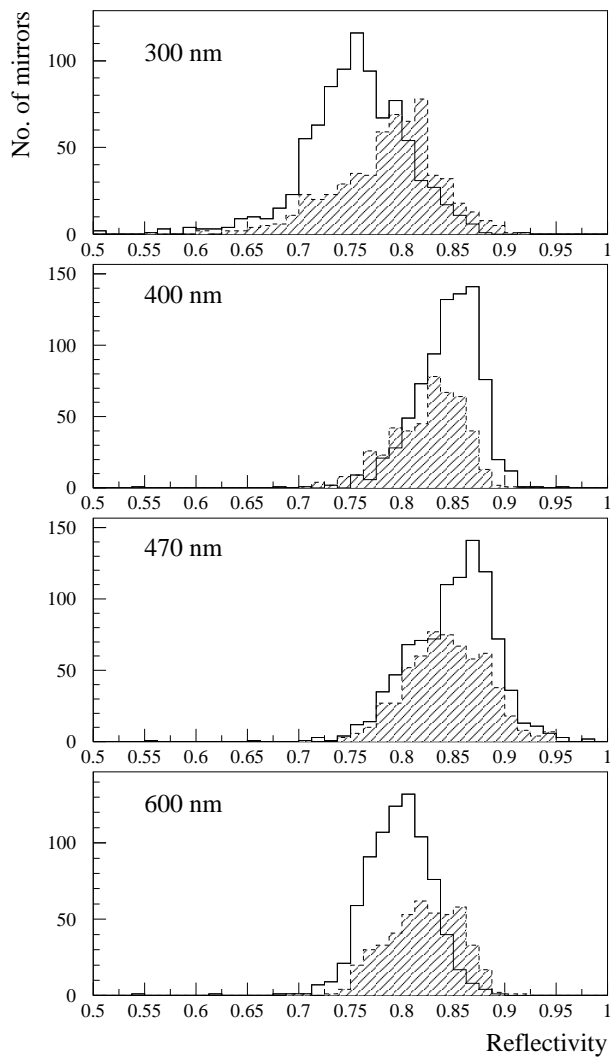}
\end{center}
\caption{Distribution of reflectivities at 300, 400, 470 and 600\,nm,
for COMPAS mirrors (line) and Galaktica mirrors (filled area). All measurements
are shown, including mirrors later rejected due to insufficient
reflectivity at one or more wavelengths.}
\label{fig_ref}
\vspace{0.5cm}
\end{figure}

Whereas the measured spot sizes gave little reason to reject mirrors,
reflectivities were a reason of concern. Figs.~\ref{fig_ref}(a) to (d)
show the reflectivities measured at the four wavelengths, for
both types of mirrors. A significant number of mirrors fall 
below the initial 80\% reflectivity criterion, in particular at 300\,nm.
In order to achieve an acceptable yield, and 
taking into account the uncertainties in the measurement, rejection
criteria were lowered to $< 70\%$ at 300\,nm and $<75 \%$ at any of
the larger wavelengths. Reflectivities measured at different
wavelengths show some correlation between neighboring wavelength bands.
Very little correlation is observed, however, between the values for
300\,nm and 600\,nm.

\begin{figure}
\begin{center}
\includegraphics[width=12.0cm]{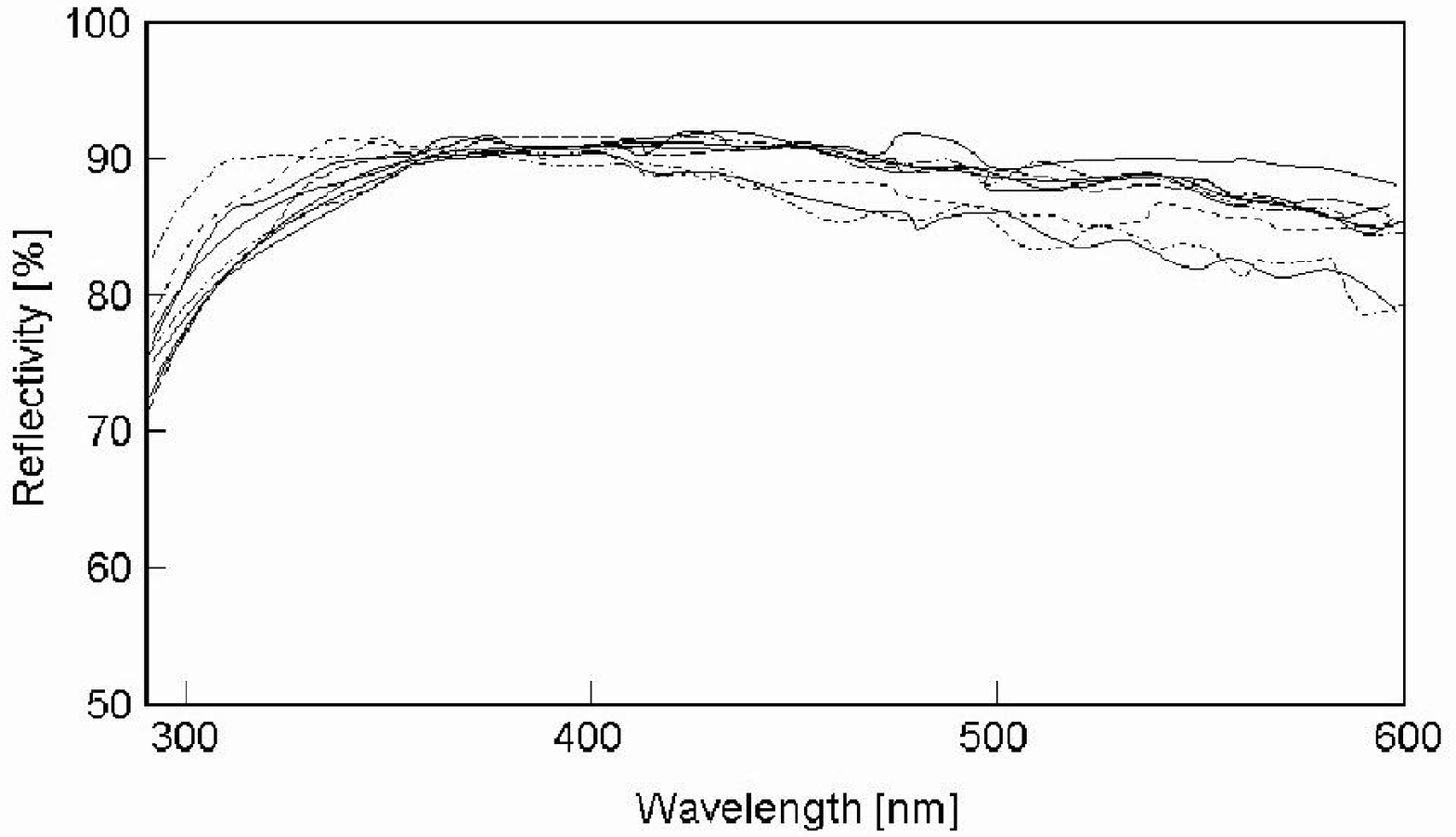}
\end{center}
\caption{Reflectivity as a function of wavelength, measured with a commercial
instrument by COMPAS, for a typical sample of mirrors.}
\label{fig_ref2}
\vspace{0.5cm}
\end{figure}

The reflectivity of the COMPAS mirrors was checked at 
the factory with a commercial
reflectometer; Fig.~\ref{fig_ref2} illustrates some of the
results for fairly typical mirrors.  The measured reflectivities
are systematically higher by about 5\%
than the typical values measured in our laboratory; the reason is not fully
understood, but may be related to the fact that in {the measurements
at the factory} only selected
spots on the mirror are measured, whereas in {our measurements} the
whole surface is averaged. Furthermore, a slightly diffuse
halo is ignored in {our} scans of the spot at $2f$, but may partly
be counted in the reflectometer.

\begin{figure}
\begin{center}
\includegraphics[width=10.0cm]{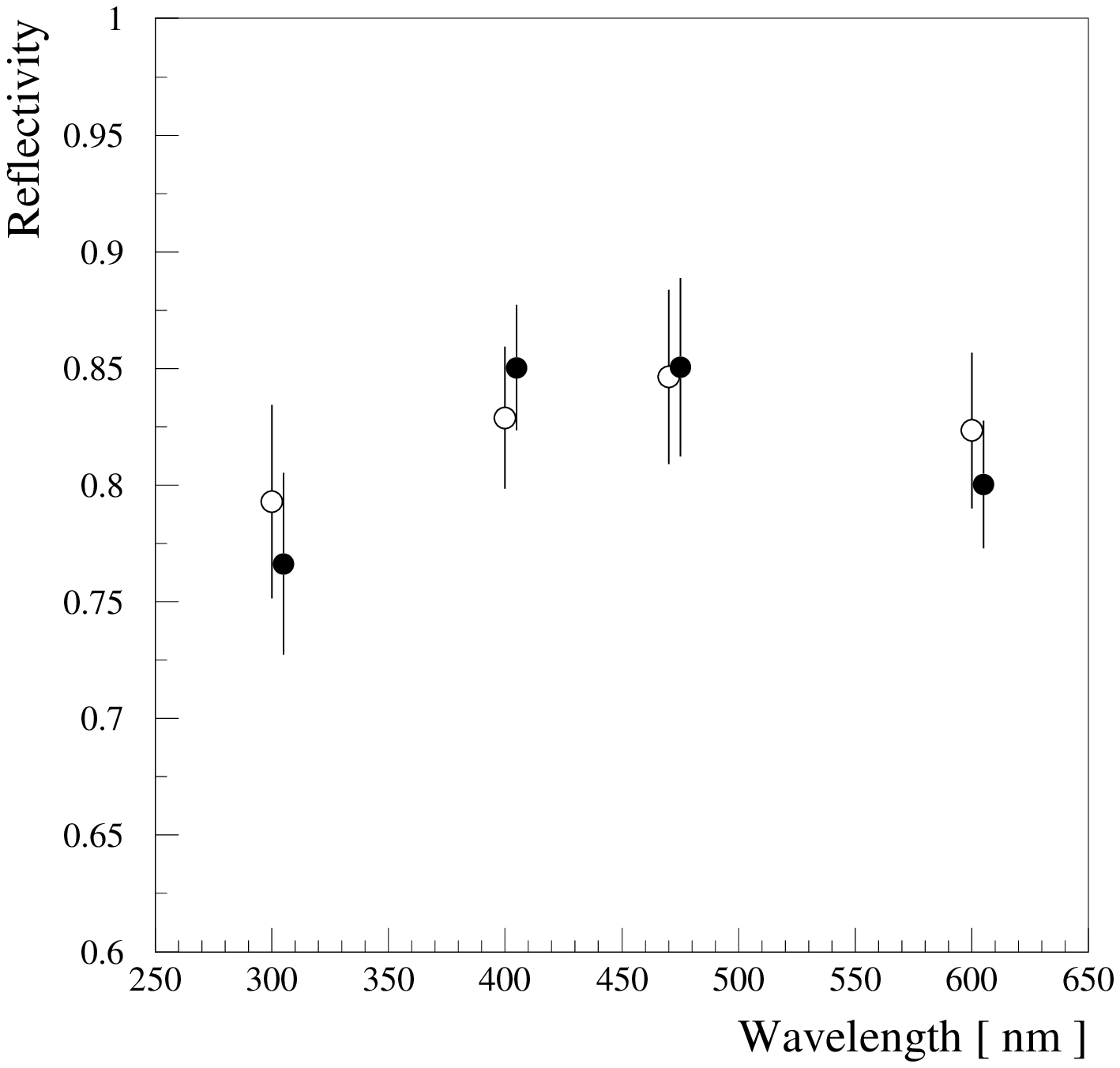}
\end{center}
\caption{Average reflectivity of accepted mirrors as a 
function of wavelength, for COMPAS mirrors (full symbols)
and Galaktica mirrors (open symbols). The vertical error
bars indicate the rms variation of reflectivities over the 
mirror samples.}
\label{fig_final_ref}
\vspace{0.5cm}
\end{figure}

Of the mirrors delivered, 4\% and 12\% of COMPAS and Galaktica mirrors,
respectively, were rejected because of mechanical defects or other
problems found in the initial inspection. All measured spot sizes were
considered acceptable. Because of insufficient
reflectivity, 20\% and 16\%, respectively, were returned for recoating
after the measurements. 
Fig.~\ref{fig_final_ref} shows the average
reflectivity of the accepted mirrors passing all tests.

\subsection{Long-term stability of mirrors}

Experience with very similar quartz coated glass mirrors has also been reported from
HEGRA \cite{puehlhofer_icrc}; there, a loss of reflectivity of 3-4\% per year is quoted,
which means that mirrors have to be recoated at least every five years.
H.E.S.S. mirrors were exposed for about two years both in Namibia 
(6 mirrors) and
in Heidelberg (2 mirrors). 
The observed deterioration was consistent with the
HEGRA results, although the number of mirrors was too small to draw a 
definitive conclusion, also given
the systematic errors in the reflectivity measurements, where over time
different instruments were employed. 

\section{Winston cones}

Most modern Cherenkov telescopes use non-imaging light concentrators
-- i.e., funnels of some type -- in front of the photomultiplier tubes.
These concentrators serve a dual purpose:
\begin{itemize}
\item They avoid dead areas due to insensitive areas
at the outer edges of the PMT cathodes and due to the support structure
in between PMTs.
\item They limit the solid angle viewed by the PMT and reduce noise due
to stray light from the ground, shining past the reflector, or from the sky
in case the telescope is observing at low elevations.
\end{itemize}
The angular acceptance of a light concentrator and the ratio of its
input and output apertures are closely related. According to Liouville's
Theorem, the phase space volume filled by light, 
$\int \int \cos \theta \mbox{d} S \mbox{d} \Omega = 
(\pi^2/4) D^2 \sin^2 \theta_{max}$ is conserved; here we have assumed uniform
illumination over a diameter $D$ and a range of angles up to $\theta_{max}$.
Winston cones \cite{winston_cone} are a special design of concentrator,
optimized for a sharp angular cutoff. For an ideal Winston cone with 
entrance diameter $D_{in}$ and exit diameter $D_{out}$, the cutoff angle
is given by $\sin  \theta_{max} = D_{out}/D_{in}$.

\subsection{Optimization and layout of the Winston cones}

\begin{figure}
\begin{center}
\includegraphics[width=7.0cm]{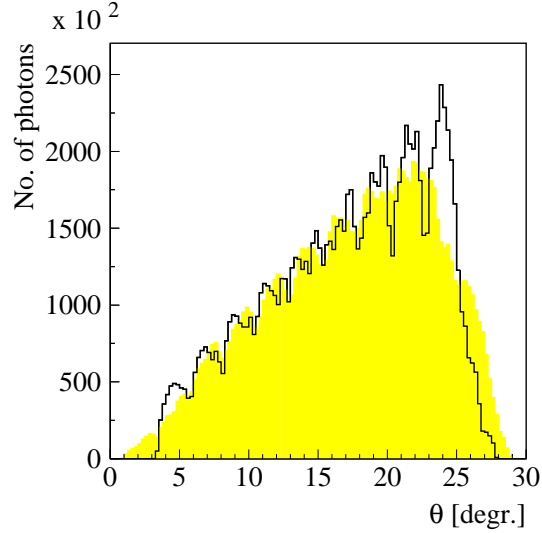}
\end{center}
\caption{Angles under which light is incident on a pixel,
for a pixel at the center of the camera (line) and 
for a pixel at the edge (filled area).}
\label{fig_dish_angles}
\vspace{0.5cm}
\end{figure}

\begin{figure}
\begin{center}
\includegraphics[width=7.0cm]{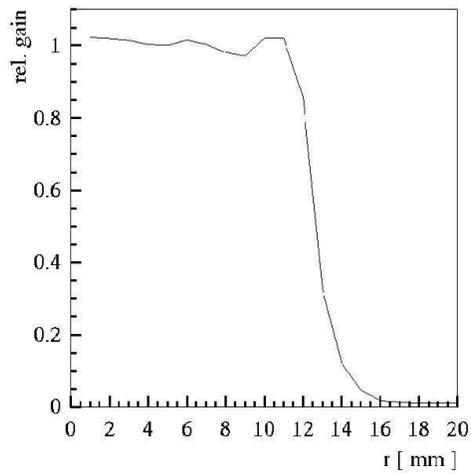}
\end{center}
\caption{PMT sensitivity as a function of the distance from the
center of the photocathode. The function has been determined from
cathode scans of 18 PMTs and is normalized to 1 over the central
21\,mm diameter circular area of the photocathodes.}
\label{fig_pmt_cathode}
\vspace{0.5cm}
\end{figure}

Boundary conditions for the optimization of the Winston cone light
concentrators of the H.E.S.S. camera were
\cite{int_winston}:
\begin{itemize}
\item the angular cutoff coincides with the outer edge of the dish
(Fig.~\ref{fig_dish_angles}),
\item the input area has a hexagonal shape in order to cover the 
focal plane with minimal losses,
\item the output area must not exceed the sensitive area of the
PMT cathode (Fig.~\ref{fig_pmt_cathode}), and
\item the cone surfaces must provide high reflectivity.
\end{itemize}

\begin{figure}
\begin{center}
\includegraphics[width=7.0cm]{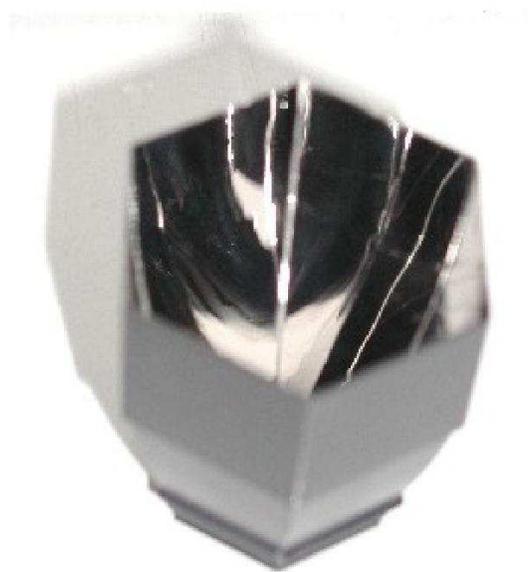}
\end{center}
\caption{One of the Winston cone light collectors.}
\label{fig_wcones}
\vspace{0.5cm}
\end{figure}

On the basis of ray-tracing simulations, a cone geometry with a
hexagonal input aperture of 41\,mm width (flat-to-flat), a 
hexagonal output aperture of 21.5\,mm width, and a length of 53\,mm
was adopted. The Winston cones (Fig.~\ref{fig_wcones}) are
made of extruded plastic, composed of two halves,
and their inner sides are aluminized and covered with a thin
quartz coating. The individual cones are locked in a carrier
plate, which defines their locations with high precision.
With a thickness of the cone walls at the entrance of the cone
of 0.5\,mm, the typical dead space between adjacent cones is
about 1\,mm, equivalent to 5\,\% of the area.

\subsection{Measurements}

\begin{figure}
\begin{center}
\includegraphics[height=8.0cm]{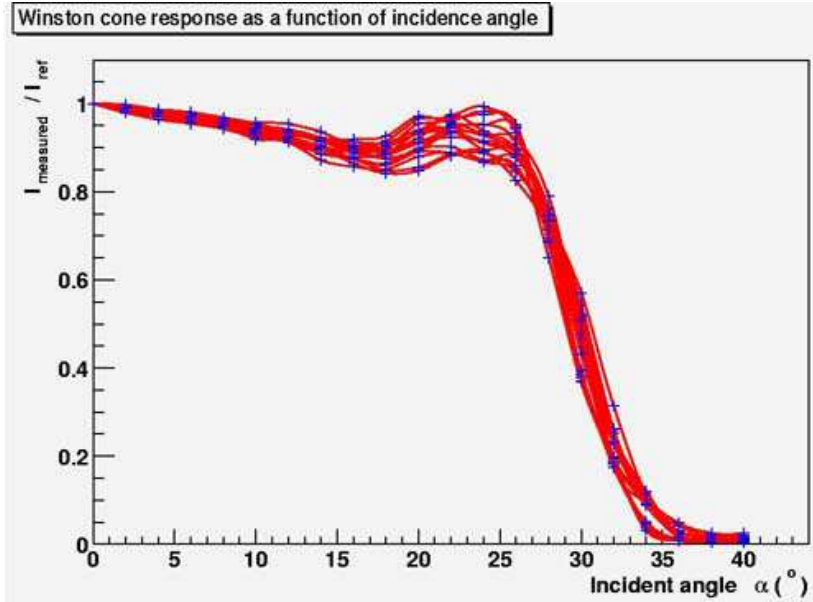}
\end{center}
\caption{Transmission of a typical Winston cone as a function of 
angle of incidence, normalized to normal incidence. Results
for several azimuthal angles $\phi$ are superimposed; $\phi$
is the rotation angle around the cone axis.}
\label{fig_wang}
\vspace{0.5cm}
\end{figure}

The transmission of all Winston cones was measured in a custom-built
setup. Fig.~\ref{fig_wang} illustrates the
transmission of a typical cone as a function of the angle of incidence. The cones
exhibit a fairly sharp cutoff at about $27^\circ$, coincident with
the angle under which the camera views the edge of the dish
(Fig.~\ref{fig_dish_angles}). The absolute transmission of all 
Winson cones was measured using a back-lit hexagonal diffusive
screen as a light source and a XP2020 PMT as photon detector; the screen covers the same range of angles
as the mirrors on the telescope. The PMT current was then determined
both with the Winston cone and with a hexagonal mask, and normalized to
the entrance area of the cone, resp. to the area of the mask.
The transmission is defined as the ratio with and without cone.
Average transmission values for the cones used in the cameras
are 0.70, 0.67, 0.79, 0.80, and 0.77
for wavelengths of 300\,nm, 350\,nm, 400\,nm, 450\,nm and 500\,nm,
respectively. The rms scatter of transmission values 
among different cones is about 0.02,
with strong correlations between different wavelengths.
Slightly over 10\% of the produced cones were rejected because of
poor transmission.

\section{Dish and mirror support units}

The full optical performance of the telescope can only be
realized if the support structure {is rigid enough that the 
relative positions and orientations of the mirrors are maintained
under operating conditions}, such as variable
elevation and wind. The H.E.S.S. telescopes rely on a stiff
steel structure to hold the mirrors, and on motorized mirror
support units to provide the fine adjustments. The mechanics is
designed such that the mirrors should stay aligned over long periods
(many months) without a need for realignment.

\subsection{The telescope dish}

The dish is constructed from steel, consisting of a shell made of
twelve strong (260 by 260 mm$^2$) 
radial beams dividing the dish into 12 sectors
(Figs.~\ref{fig_telescope},\ref{fig_mirr_arr}). In each
sector, 10 mirror support tubes (159\,mm diameter) carry mounting
brackets to attach the mirrors, via their mirror support units.
A spaceframe made of 83 mm diameter steel tubes provides depth
and stiffness for the structure. 
Additional stiffness is provided by tensioning the dish along
its elevation axis with a force of approximately 100\,kN, 
applied via the two support
towers and the baseframe. The total mass of dish structure is 
23.4\,t, including the mirror mounting brackets. 
The mirror support units and mirrors add 7.6\,t; 
additional masses include 3.25\,t of the camera support, 2.46\,t of the
elevation drive rail and 0.8\,t of 
counterweights required to balance the dish.

The telescope structure was designed by SBP\footnote{Schlaich Bergermann und Partner,
Consulting Engineers, Stuttgart} and produced by 
NEC-Stahl\footnote{Namibia Engineering Corporation, NEC-Stahl, Okahandija}
based on production drawings by SCE\footnote{Seelenbinder Consulting
Engineers, Windhoek}.

We note that the spherical shell required for an exact Davies-Cotton
layout is approximated by straight mirror support tubes in each of the
twelve dish sectors. The geometry of the radial support beams, with a
curvature radius of 15.86\,m, was optimized to minimize the deviations
from an exact Davies-Cotton layout. In the worst cases, mirrors are
displaced by about 8\,cm in the $z$ direction (parallel to the optical
axis) from their nominal positions. Such a displacement is tolerable
due to the relatively small size of the mirror facets; if the focal
plane is displaced by 15\,cm, the spot of an otherwise ideal mirror of
60\,cm diameter will grow to 0.6\,cm diameter, or 0.4\,mrad, well below
the pixel size.

\begin{figure}
\begin{center}
\includegraphics[width=9.0cm]{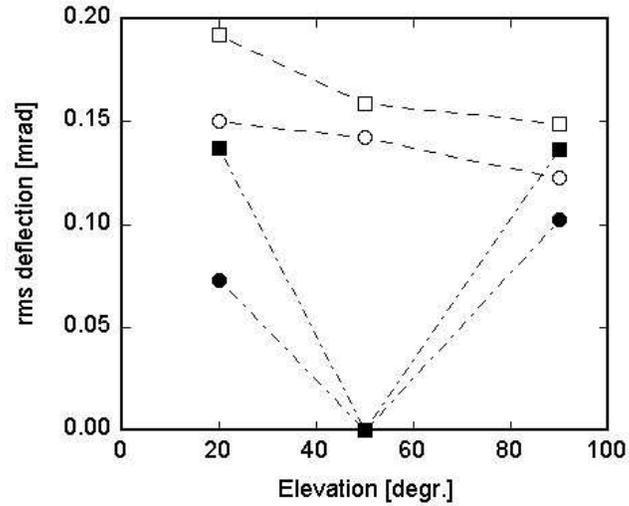}
\end{center}
\caption{Rms deflection 
of mirrors in the horizontal (squares) and vertical (circles) directions
as a function of elevation, according to FEM simulations. Open symbols show the total deflection
under gravity, closed symbols the deflection relative to 
an assumed alignment position
at $50^\circ$.}
\label{fig_dishdef}
\vspace{0.5cm}
\end{figure}

Extensive FEM simulations of the dish were carried out by SBP to evaluate
and optimize the stability of the dish, under gravity load, wind loads,
and under the influence of deformations of the baseframe induced by slight
unevenness of the azimuthal rail. The
predicted gravity-induced
deformations of the dish are illustrated
in Fig.~\ref{fig_dishdef}. Two sets of points are given, one corresponding to
the total deflection of mirrors under the influence of gravity at various
elevations, and one assuming that mirror facets are aligned at $50^\circ$ elevation,
such that only the deflection relative to the alignment position matters.
Compared to these gravity loads, the influence of wind loads etc. is small.

The main specifications of mount and dish are summarized in
Table~2.

\begin{table}
\begin{tabular}{|l|l|}
\hline
\multicolumn{2}{|l|}{Dish} \\
\hline
Number and type of mirrors & about 380 60\,cm mirrors \\
& weight incl. support $\approx$ 20\,kg each \\
Mirror spacing & 62\,cm center-to-center \\
Mechanical tolerances of dish structure & 5\,mm \\
Tolerance on orientation of mirror mounting brackets & $0.5^\circ$ \\
Variation of mirror orientation with elevation & $< 0.14$\,mrad rms in each projection\\
& for $20^\circ$ to $90^\circ$ elevation\\
\hline
\multicolumn{2}{|l|}{Mount} \\
\hline
Turning range in azimuth of mount & $\ge 385^\circ$ \\
Elevation range of mount & $-35^\circ$ to $+175^\circ$\\
Slewing speed in azimuth and elevation & $100^\circ$/min \\
Positioning accuracy & $0.01^\circ$ \\
Max. wind speed during operation & 50\,km/h \\
Max. wind speed & 160\,km/h \\
\hline
\end{tabular}
\label{dish_specs}
\vspace{0.3cm}
\caption{Initial specifications for mount and dish.}
\vspace{0.5cm}
\end{table}

A final point of discussion was
the coloring scheme of the mount and dish structure. There are
two conflicting requirements. To reduce heating of the structure during
daytime, it is best painted white. On the other hand, 
for nighttime observations a dark color is preferred, since it (a)
minimizes stray light -- part of the telescope structure is visible
from the camera, e.g. in between mirrors -- and since it (b) helps
to reach thermal equilibrium more quickly. Most of the insolation
during daytime is at long wavelengths, in the red and IR, whereas the
PMTs of the camera are sensitive mainly to blue light. 
A red color provides a good compromise, since 
it reflects at long wavelengths and absorbs in the blue. Characteristics
of different colors were studied by measuring the temperature increase
when structures were exposed to sunlight, and by evaluating the reflectivity
as a function of wavelength \cite{color}.
Following objective criteria, a {bright red/pink color} would be optimal. As a compromise
between these criteria and the subjective impression, a slightly darker
red tone was adopted.

\subsection{The mirror support units}

{Purpose of} the mirror support units is (a) to attach the mirrors
firmly to the dish structure, and (b) to {enable} the remote adjustment of mirrors. In order
to prevent {damage to, or deformations of} the mirrors, the mirror support
system must not impose significant stress onto the mirror, and must allow for
differential thermal expansion of the mirror and its support.
The key specifications for the mirror support units are 
summarized in Table~3.

\begin{table}
\begin{tabular}{|l|l|}
\hline
\multicolumn{2}{|l|}{\bf Mirror support units} \\
\hline
Load & Mirror weight 10 - 15 kp, wind load negligible during normal operation \\
     & Peak wind load 30 kp \\
Support & Must not overconstrain the mirror, or induce stress \\
Stability & Mirror orientation stable to 0.1\,mrad (rms) under operating
   conditions \\
Adjustment range & Mirror orientation adjustable within $\pm1^\circ$ 
($\pm$ 17\,mrad)\\
Adjustment precision & 0.1 mrad maximum deviation, relative to last position; \\
 & only relative positioning required \\
Speed & at least $1^\circ$/min.\\
Environmental conditions & Temperatures $-10^\circ$ to $60^\circ$ 
($0^\circ$ to $30^\circ$ during adjustments) \\
& Humidity 0\% to 100\% ; high UV radiation; dust and sand \\ 
\hline  
\end{tabular}
\vspace{0.3cm}
\caption{Initial specifications for mirror support units.}
\label{support_specs}
\vspace{0.5cm}
\end{table}

\begin{figure}
\begin{center}
\includegraphics[width=10.0cm]{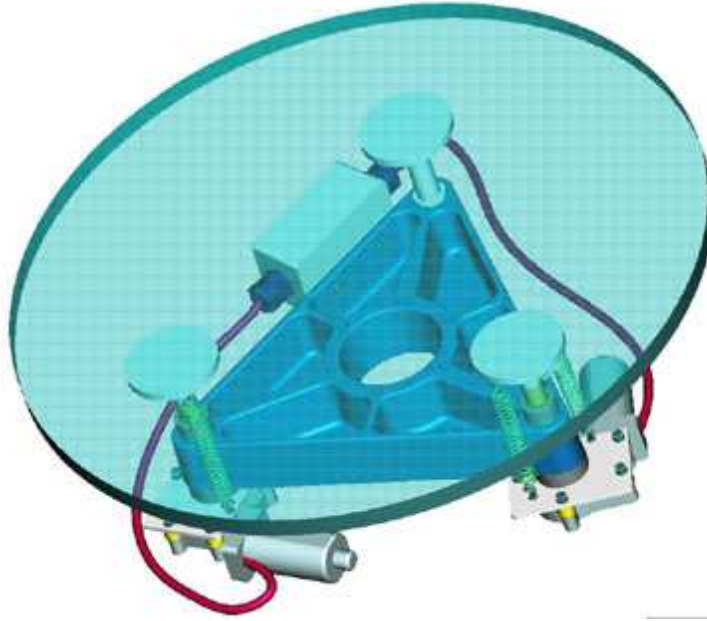}
\end{center}
\caption{Mirror support unit. 
The unit includes two motorized actuators and a
connector box.}
\label{fig_mirror_supp}
\vspace{0.5cm}
\end{figure}

The solution adopted is illustrated in Fig.~\ref{fig_mirror_supp}.
A mirror unit consists of a support triangle carrying one fixed
mirror support point and two motor-driven actuators. The mirror is
attached at these three points using steel pads glued to the back of
the mirror. The whole unit is preassembled and then with three screws
attached to the mounting bracket welded to the mirror support tube of 
the dish. The mounting brackets have a machined surface and are pre-aligned before the welding,
to a precision of $0.5^\circ$, so that the support unit provides only
the final alignment within a limited range.

The basic support triangle underwent several design iterations, until
finally a 5 cm thick cast-aluminum structure proved to be stable enough. Under
the load of the mirror, the triangle (and the mounting bracket) is
deformed by about 6\,$\mu$m, equivalent to an angular deflection of
0.02\,mrad. 

The actuators are based on automotive products used to drive car
windows\footnote{Bosch, Type 0 130 821 708/709}. 
A motor unit includes the drive motor, two Hall sensors
shifted by $90^\circ$ sensing the motor revolutions and providing
four TTL signals per turn, and a 55:1 worm gear with a nominal speed
at the exit of 100\,rpm and a torque of 1.5\,Nm, increasing to 6\,Nm at
lower speeds. The motor is directly coupled to a 12 mm threaded bolt,
driving the actuator shaft by 0.75\,mm per revolution. The total range
of an actuator is 30\,mm; one count of the Hall sensor corresponds to
a step size of 3.4\,$\mu$m, or about 0.013\,mrad tilt of the mirror.

The mirror is supported in three points, using 8\,cm diameter steel pads
glued to the back of the mirror. To reduce stress, a flexible 3\,mm glue
layer\footnote{TEROSTAT 8590 (black)} is used; a 
primer\footnote{TEROSTAT PPRIMER 8511} provides optimum contact
to the glass and a sealant\footnote{TEROSTAT DICHTSTOFF M935 (white)} around the circumference
protects the glue from UV solar radiation. 

\begin{figure}
\begin{center}
\includegraphics[width=12.0cm]{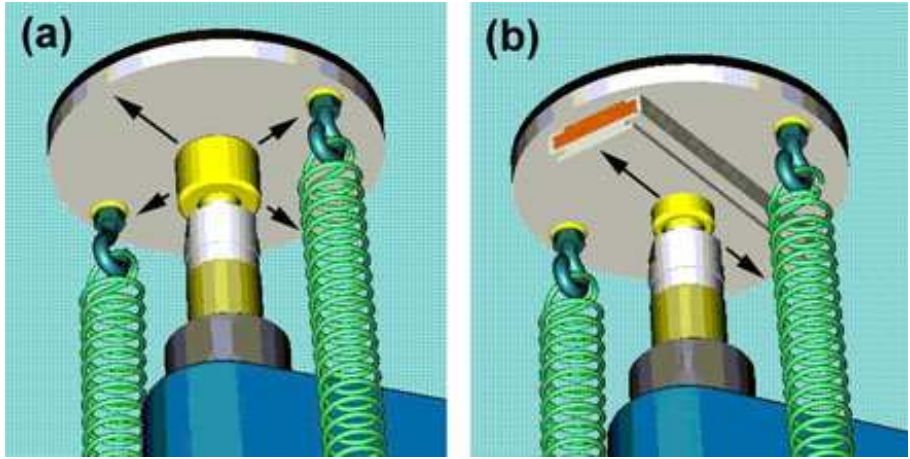}
\end{center}
\caption{The actuator heads and their connection
to the mirror. One head (a) is free to slide in both directions
on a steel pad glued to the mirror, the other (b) can move in a 
linear sliding mechanism. The third support point ends in a 
spherical joint, allowing only rotational degrees of freedom.}
\label{fig_head}
\vspace{0.5cm}
\end{figure}

Of the three support points, one -- the fixed point -- is equipped
with a spherical joint. One of the actuator heads 
(Fig.~\ref{fig_head}(b)) is attached to the steel
pad via a spherical joint and a one-dimensional 
slider\footnote{IGUS Drylin N}.
The second head (Fig.~\ref{fig_head}(a)) 
can slide in two directions on the steel pad, with a plastic pad
\footnote{IGUS iglidur G}
added to reduce friction. 
Springs at each support point pull the steel pads towards the support,
hold the mirror in place and eliminate play. This
arrangement, where all support points can adjust their orientation, one
is fixed, one moves in one direction and one moves in two directions in
the plane of the mirror, fully constrains the position and orientation 
of the mirror, yet allows the mirror to expand and avoids stress on 
the mirror when moved.

The performance and reliability of the mirror support units was tested
extensively by cycling the temperature over extended periods
between
$-10^\circ$ and $60^\circ$, and the humidity between 10\% and 80\%.
At the same time, the units where exposed to UV radiation from a 
1 kW metal halid lamp, simulating the conditions in Namibia. Of particular
concern was the behavior of the thick glue layer; it shows a constant, linear
$0.50 \pm 0.02\,\mu$m/$^\circ$C expansion, which should not influence the
mirror pointing. We verified that image spots do not vary
significantly when a mirror is moved, which would be a sign of stress-induced
deformations.

\subsection{The mirror control system}

\begin{figure}
\begin{center}
\includegraphics[width=8.0cm]{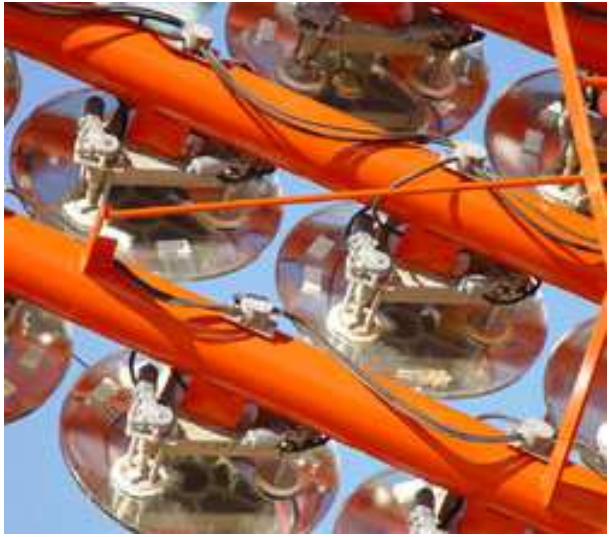}
\end{center}
\caption{Section of the dish showing mirrors,
the mirror mounting brackets, the branch
cable and the short connector cables.}
\label{fig_branch_photo}
\vspace{0.5cm}
\end{figure}

A special control system was designed to control all actuator motors of a
telescope \cite{mirror_ctrl}.
The dish is divided into twelve segments with up to 32 mirror facets
(64 motors) each, and one branch cable runs along the mirrors of
one segment to connect them to the control system.
For each mirror, a short cable branches off the main branch cable and is
plugged into a small relay box attached to the mirror support triangle
(see Fig.~\ref{fig_branch_photo}).
Within a branch, one motor out of 64 is selected using 2~$\times$~8 wires which
form an address matrix, implying that only one motor can be driven
at a time.
The twelve branch cables of a telescope are connected to branch driver boards
located in an electronics hut on the baseframe.
The branch driver boards provide all necessary power and signals, and are
controlled via a VME bus interface.
The Hall signals of the selected motor are fed into a special decoder
\footnote{Agilent, HCTL-2020} which
derives information about the direction of movement and the number of Hall
counts, i.e. the distance of actuator travel.
It implements a set of signal filters and consistency checks to avoid
miscounts.
Note that the Hall sensors do not {allow one} to determine the absolute position of an
actuator.
This restriction can be addressed by using one of the actuator stops as an
internal position reference, and by using the control
system to turn off the motor after a predetermined
number of counts, so that a precise relative positioning is achieved.
In order to position an actuator, usually 2-3 iterations are required, since
-- depending on its speed and load -- the motor may keep turning for a few Hall
counts after the drive voltage is switched off.
To simplify the system, the actuators are not equipped with end switches.
Instead, springs stop the actuator at the ends of its range, and the control
system detects the missing Hall counts and shuts off the actuator
motor.

\subsection{Tests of the mirror alignment system}

\begin{figure}
\begin{center}
\includegraphics[width=8.0cm]{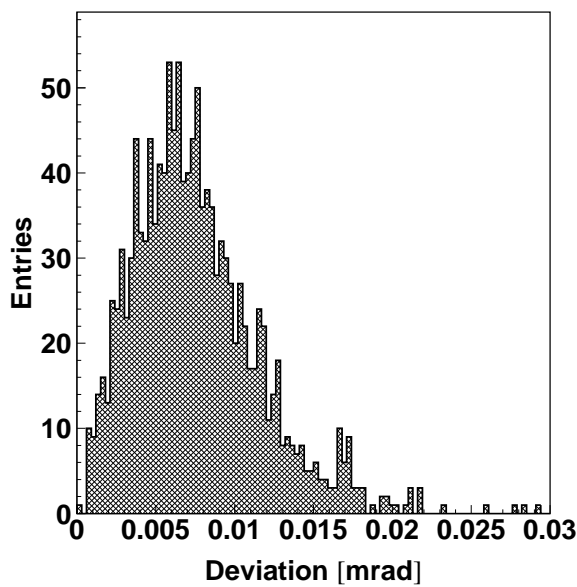}
\end{center}
\caption{Deviation between initial and final mirror pointing for test
cycles, where first actuator 1 was moved over 6\,mm, then actuator 2 over
6\,mm, followed by moving actuator 1 over -6\,mm and actuator 2 over -6\,mm,
bringing the mirror back to its starting point. The rms deviation is
0.0086\,mrad.}
\label{fig_dev}
\vspace{0.5cm}
\end{figure}

The actuators and control systems were tested extensively.
Tests of the control electronics showed that no miscounts of Hall signals occur,
implying that the (iterative) positioning of actuators can be performed with a
precision of one Hall count, or about 3$\,\mu$m \cite{cornils_dipl}.
In tests where the reflected spot of a mirror was monitored by a CCD camera, 
a rms positioning accuracy of the mirror of 0.0086\,mrad
was achieved consistently (Fig.~\ref{fig_dev}); 
this value is only slightly larger than the expected optimum
of 0.0075\,mrad,
based on the finite step size of the actuator,
and is absolutely minute compared to the mirror spot size.

\section{Effective reflector area}

An essential performance characteristic of the telescope and input
for simulations is the 
effective reflector area. It is governed by the following factors
\cite{shadow}:
\begin{itemize}
\item the number of mirror facets on a telescope is 380 
(the dish can hold 382; two mirrors are omitted to
provide space for the optical guide telescope),
\item {in particular the outer mirror facets are inclined 
(by up to $13^\circ$)
with respect to the plane of incident light, reducing the effective 
reflector area by 1.1\% in comparison to 380 times the 
facet area}, 
\item the camera arms and their bracing struts shadow about 5-6\% of 
the incident light, and
\item some of the bracing struts in addition shadow reflected light,
causing an additional 5-6\% reduction in light yield.
\end{itemize}

\begin{figure}
\begin{center}
\includegraphics[width=11.0cm]{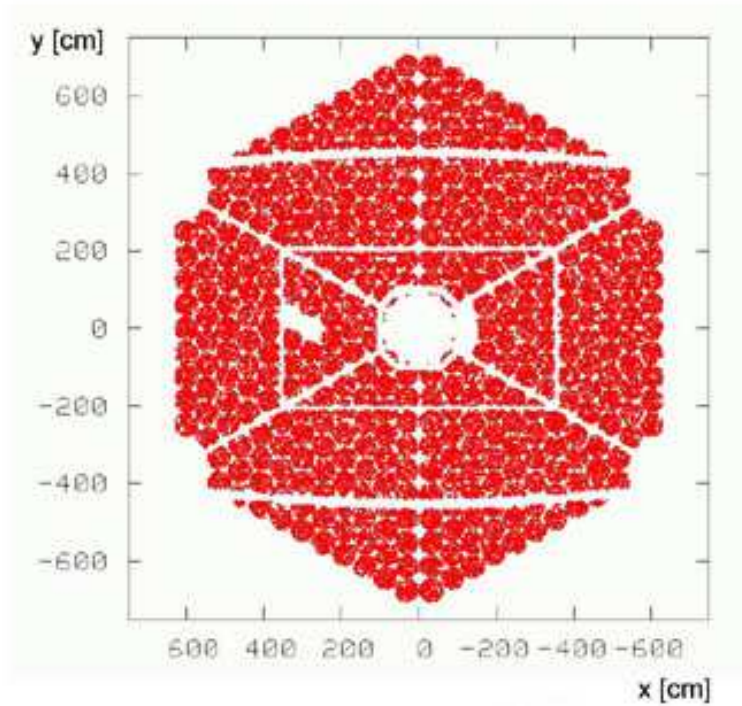}
\end{center}
\caption{Impact points on the reflector of those photons incident parallel to the optical
axis, which reach the camera after reflection on the mirror. 
The straight shadow
lines are caused by shadowing of incident light by the camera support structure;
the two slightly curved shadows correspond to light obscured on the way from
the reflector to the camera. The shadows of the diagonal bracing rods are not
visible at this resolution.}
\label{fig_shadow}
\vspace{0.5cm}
\end{figure}

\begin{figure}
\begin{center}
\includegraphics[width=9.0cm]{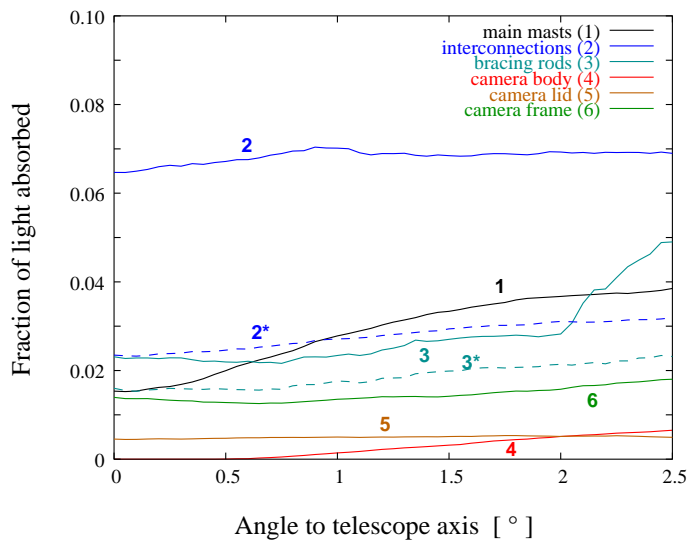}
\end{center}
\caption{Fraction of light obscured as a function of the angle to the telescope axis,
for three different azimuth angles. The field of view of the PMT cameras extends to
$2.5^\circ$.}
\label{fig_fraction}
\vspace{0.5cm}
\end{figure}

Fig.~\ref{fig_shadow} shows the results of a ray-tracing simulation
illustrating which parts of the reflector are shadowed. The exact 
fraction of light lost depends somewhat on the angle of incidence relative
to the optical axis, and also on the azimuth, and varies between 
10\% on-axis and about 12\% at the edge of the field of view, see
Fig.~\ref{fig_fraction}. The resulting effective reflector area of
a H.E.S.S. telescope varies over the field of view
between 93.4\,m$^2$ and 95.3\,m$^2$


\section{Summary}

The H.E.S.S. optical system is based on a segmented reflector with 
$f/d \approx 1.2$ and a reflector area of 107\,m$^2$; the 380 spherical
mirror facets have identical focal lengths of 15\,m and are arranged 
in a Davies-Cotton layout. Shadowing due to the camera arms 
reduces the effective reflector area to about 95\,m$^2$, depending
slightly on the angle of incidence. The mirror facets are ground glass mirrors,
aluminized and quartz coated. The mirror facets were measured individually
and typically concentrate 80\% of the reflected light within a circle
of about 0.4\,mrad diameter, well below specifications. The average reflectivity
in the relevant wavelength region is 80\% to 85\%. Including optical
aberrations, the system should provide a point spread function which
is on the optical axis significantly smaller than the pixel size of $0.16^\circ$,
and which is comparable to the pixel size near the edge of the field
of view of the PMT camera. Winston cone light concentrators added in
front of the PMT pixels serve both to improve light collection, and
to limit the field of view to the size of the reflector
to reduce albedo from the
ground. 

The mirror facets are carried by motorized support units, to allow
remote adjustments of the mirror orientation. In the design of the
dish,  mechanical stiffness and minimal deformations under gravity
load were emphasized to maintain a good point spread function
over a wide range of elevations.

The mirror alignment and the resulting measured point spread functions
are covered in a companion paper.

\section*{Acknowledgements}

A large number of persons have, at the technical level, contributed to
the design, construction, and commissioning of the telescopes. 
We would like to express
our thanks to the technicians and engineers from the participating
institutes for their devoted engagement, both at home and in the 
field, initially frequently under adverse conditions. The team on site,
T.~Hanke, E.~Tjingaete and M.~Kandjii provided excellent support. 
S. Cranz, the owner of the
farm Goellschau, has given valuable technical assistance.
We gratefully acknowledge the contributions of U.~Dillmann, 
T.~Keck, W.~Schiel of SBP, of H.~Poller of SCE and of R.~Schmidt
and F.~van~Gruenen of NEC, responsible for the design and the
construction of the telescope structures. Telescope construction
was supported by the German Ministry for Education and Research BMBF.

\end{document}